\documentclass{PoS}
\newcommand{\be}{\begin{eqnarray}}
\newcommand{\ee}{\end{eqnarray}}

\newcommand{\nn}{\nonumber }

\title{The sign problem in the $\epsilon$-regime of QCD}

\ShortTitle{The sign problem in the $\epsilon$-regime of QCD}

\author{K. Splittorff\\
        The Niels Bohr Institute, Blegdamsvej 17, DK-2100, Copenhagen {\O}, Denmark\\
        E-mail: \email{split@nbi.dk}}

\abstract{QCD in the $\epsilon$-regime at nonzero 
baryon chemical potential $\mu$ is reviewed. The focus is on aspects 
of the sign problem which are relevant for lattice QCD.    
It is discussed how spontaneous chiral symmetry breaking 
and the sign problem are related through the spectrum of the 
Dirac operator. The strength of the sign problem is 
linked to the quark mass and the chemical potential. 
Specific implications for lattice QCD are discussed.}

\FullConference{XXIVth International Symposium on Lattice Field Theory\\
                July 23-28, 2006\\
                Tucson, Arizona, USA}

\begin{document}

\section{Introduction}

A way to understand lattice QCD at nonzero chemical potential, and the
physics behind it, is to analyze the statistical properties of the Dirac
eigenvalues. Twenty years ago \cite{Barbour,Gibbs} it was realized 
that this approach apparently leads to a contradiction: At zero temperature
a small (in units of the nucleon mass) chemical potential is expected 
to have a small effect on the chiral condensate, $\Sigma$. On the contrary the 
scatter plots of the Dirac spectrum obtained in lattice simulations 
\cite{Barbour,latticespectrum} leads one to conclude that $\Sigma$ 
will vanish in the chiral 
limit for {\sl any} nonzero value of the chemical potential. This conclusion
was reached by means of an electrostatic analogy \cite{Barbour} where the 
chiral condensate is regarded as the electric field created by charges 
located at the position of the eigenvalues in the complex plane. The quark
mass serves as a test charge inserted at the point $(m,0)$ in the complex 
plane where the electric field is measured. At zero chemical potential this 
analogy correctly leads to the Banks-Casher relation \cite{BC}. For a
nonzero  
chemical potential the Dirac operator loses it anti-hermeticity and,  
consequently, the eigenvalues spread out into the complex eigenvalue plane. 
In the chiral limit the test charge (quark mass) moves to the center of this 
charge distribution and the chiral condensate hence vanishes for any nonzero 
chemical potential. The only way out seemed to be that a finite fraction 
of the eigenvalues would remain in a singular distribution (delta-function) 
on the imaginary axis at nonzero chemical potential. 
However, in unquenched QCD the sign problem creates a loophole in the
electrostatic  
analogy. The unquenched spectral density of the QCD Dirac operator 
(the charge distribution) is a complex function 
which for for $\mu>m_\pi/2$ depends strongly on the quark mass (the test
charge).  
Analytic calculations in the $\epsilon$-regime \cite{O} show that the
complex valued spectral density  has oscillations with a period inversely
proportional to 
the volume and an amplitude that grows exponentially with the volume 
\cite{AOSV}. These oscillations lead to the discontinuity of the
chiral condensate at zero quark mass \cite{OSV}. Spontaneous chiral symmetry
breaking is therefore intimately related to the sign problem. 
Because the oscillations of the eigenvalue density take place on a scale 
inversely proportional to the volume, the microscopic scaling of the
$\epsilon$-regime is required to resolve the individual oscillations
and hence the way in which the discontinuity of the chiral condensate
is built up form the eigenvalue density. 

In quenched QCD there is of course no sign problem and the electrostatic
analogy is in fact valid. By solving a random matrix model for QCD at the 
mean field level it was argued in \cite{misha} that quenched QCD at nonzero 
chemical potential is the zero flavor limit of QCD with nonzero isospin 
chemical potential. Indeed, the chiral condensate vanishes in the chiral 
limit for any nonzero isospin chemical potential \cite{eff,KTV,KogutS}.  
Lattice simulations \cite{AW,OW,BW} have already been successfully compared 
to the quenched microscopic eigenvalue density computed in \cite{SplitVerb2}.

The complex oscillations of the unquenched eigenvalue density is a
manifestation of the sign problem. The oscillations appear for $\mu>m_\pi/2$. 
In the $\epsilon$-regime it is also possible to compute the unquenched 
average of the phase factor of the fermion 
determinants and thereby to measure the strength of the sign problem 
directly \cite{phase}. It is shown that the average phase factor goes to zero 
at $\mu=m_\pi/2$ and remains at zero for 
$\mu>m_\pi/2$. The sign problem is therefore particularly acute 
for $\mu>m_\pi/2$. It is not surprising that the sign problem sets 
in at $\mu=m_\pi/2$. Suppose we had neglected the sign problem, 
that is, replaced the fermion determinant in the partition function 
by its absolute value. Since conjugating a fermion determinant 
corresponds to changing the sign of the chemical potential 
\cite{AKW} the free energy would have a second order discontinuity 
at $\mu=m_\pi/2$ signaling the formation of a Bose condensate of pions
\cite{eff}. 
Reinserting the phase factor of the fermion determinant must wipe out this 
Bose condensate completely and hence the complex nature of the fermion 
determinant must be particularly important for $\mu>m_\pi/2$. Since we 
know the physical origin of the scale $\mu=m_\pi/2$ we 
can understand the behavior of the sign problem at nonzero 
temperature and chemical potential. This allows us to make  
contact to lattice simulations \cite{FK,BieSwan,GG,DeFPh,Lombardo} at 
nonzero temperature and chemical potential.

This review is organized as follows. We first define the $\epsilon$-regime
and briefly discuss two independent ways to compute the microscopic
spectral correlation functions. Then we analyze the unquenched eigenvalue 
density in detail. In section \ref{sec:BC} we show how the oscillations of
the eigenvalue density lead to the discontinuity of the chiral condensate.
Finally we discuss the strength of the sign problem and the consequences for
lattice simulations.

\section{The $\epsilon$-regime at nonzero chemical potential}

In order to extend the Banks-Casher relation to nonzero chemical
potential it would be helpful to know the eigenvalue density of the Dirac
operator near the origin, in the phase where chiral 
symmetry is spontaneously broken, for small $\mu$ and $m$. 
Remarkably, in the $\epsilon$-regime it is possible to get all 
this. By definition \cite{GLeps} the $\epsilon$-regime deals with the phase
where  
chiral symmetry is spontaneously broken. The quark mass and chemical
potential are taken such that ($V$ is the 4-volume and $F_\pi$ is the pion
decay constant)
\be        
m\Sigma\sim\frac{1}{V} \ \ 
{\rm and} \ \ \ \mu^2 F_\pi^2 \sim \frac{1}{V} .
\label{scaling}
\ee
The correlations of the Dirac eigenvalues, $z$, are considered on the
microscopic scale \cite{SV} where 
\be
z\Sigma\sim\frac{1}{V}.
\ee

The original work on the $\epsilon$-regime \cite{GLeps} focused on the quark
mass dependence of the finite volume partition function and shows how the 
chiral condensate goes to zero if the quark mass is taken to zero in a finite
but large volume.    
The effect of the chemical potential on the finite volume partition function 
is determined by the flavor symmetries and the scaling of the chemical potential
with the volume. Using the GOR relation and (\ref{scaling}) it follows that 
the chemical potential, in the $\epsilon$-regime, is of the same order as the
pion mass. With this scaling the Compton wavelength of the pion is much
larger than the linear size of the volume and the effective partition
function reduces to a group integral over the static modes of the pion field uniquely 
determined by the pattern of chiral symmetry breaking \cite{TV}
\be
Z_{N_f}(\{m_f\};\mu) = 
\int_{U \in U(N_f)} dU \ \det(U)^\nu\ \mbox{e}^{-\frac {V}{4}F_\pi^2\mu^2
{\rm Tr} [U,B][U^{-1},B]\ +\ \frac 12 \Sigma V {\rm Tr}M(U + U^{-1})}.
\label{zeff}
\ee
Both $M$ and $B$ are diagonal matrices. $M$ is the quark mass matrix and $B$
contains the quark baryon charges. Since here all quarks have the same baryon
charge the $B$ matrix is proportional to the unit matrix and the dependence
on the chemical potential automatically drops out of the partition
function (\ref{zeff}). This is exactly as expected, since the pions have
baryon charge zero, the chemical potential is inert.

\subsection{The eigenvalue density}

The zero temperature effective partition function does not depend 
on the chemical potential even though the eigenvalues of the Dirac operator 
do\footnote{To understand how this is possible is frequently referred to as
  {\sl the silver blaze problem} \cite{cohen}.}.  
The chemical potential adds a hermitian part to the 
Dirac operator and the eigenvalue spectrum, 
\be
(D+\mu\gamma_0)\psi_j=z_j\psi_j,
\ee 
is no longer purely imaginary. The support of the eigenvalue density 
\be
\label{defdens}
\rho_{N_f}(z,z^*,\{m_f\};\mu)
\equiv \left\langle \sum_{j}\delta^2(z-z_j)\right \rangle_{N_f},
\ee
is therefore two dimensional domain in the complex eigenvalue plane. 
Here we have used the notation 
\be
\left\langle{\cal O}\right\rangle_{N_f}\equiv\frac{\int{\rm
    d}A\ {\cal O}\ \prod_{f=1}^{N_f}\det(D +
  \mu\gamma_0+m_f) \ \mbox{e}^{-S_{\rm YM}}}{\int{\rm
    d}A \prod_{f=1}^{N_f}\det(D +
  \mu\gamma_0+m_f) \ \mbox{e}^{-S_{\rm YM}}} .
\ee 
The eigenvalue density is the function which allows us to turn the average 
of a sum over eigenvalues, e.g.~$\langle \sum 1/(z_j+m)\rangle$, into an
integral,  
$\int d^2z \; \rho/(z_j+m)$. However, due to the presence of the complex
fermion determinant in the measure (the sign problem) the unquenched 
eigenvalue density is not expected to be real and positive.

The eigenvalue density in the $\epsilon$-regime, also known as the 
microscopic eigenvalue density, describes the 
eigenvalues in a range of order $1/\Sigma V$ from the origin. 
At present there exist two independent ways to compute the microscopic
correlation functions of the QCD Dirac operator at nonzero baryon chemical
potential. One can obtain them directly from the effective partition
functions \cite{SplitVerb2,AOSV} writing the $\delta$-functions in
(\ref{defdens}) by means of the replica trick \cite{misha,SplitVerb1} 
\be
\!\!\!\rho_{N_f}(z,z^*,\{m_f\};\mu)= 
\frac{1}{Z_{N_f}}\lim_{n\to0}\frac{1}{n} \partial_z \partial_{z^*}\!\! 
\int{\rm d}A |\det(D + \mu\gamma_0+z)|^{2n}
\prod_{f=1}^{N_f}\det(D + \mu\gamma_0+m_f) \ 
\mbox{e}^{-S_{\rm YM}}.
\label{replicadens}
\ee
Note that the integral is a partition function with $n$ additional quarks and
conjugate quarks. The conjugate quarks corresponds to quarks with the
opposite baryon charge \cite{AKW} and this is why these partition functions and
hence the eigenvalue density depend on $\mu$. We refer to these partition
functions as the generating functionals for the eigenvalue density.

Alternatively one can start from a chiral random matrix theory and use 
biorthogonal polynomials in the complex plane \cite{AV,A03,B,O}. 
For an up-to-date review of the random matrix approach to QCD at nonzero
chemical potential, see \cite{gernot-review}. 
(In principle, one can also make use of the supersymmetric method
\cite{Efetov}.)

The microscopic eigenvalue density of the Dirac operator in quenched QCD 
was first obtained from effective partition functions like (\ref{zeff}) 
in \cite{SplitVerb2} and
subsequently reproduced from the random matrix methods \cite{O}. 
The expression for the quenched eigenvalue density is 
\be\label{rhoquenched}
\rho_{N_f=0}(z,z^*;\mu)
 &=&  \frac{|z|^2\Sigma^4V^3}{2\pi\mu^2F_\pi^2}\mbox{e}^{-2\mu^2 F_\pi^2 V}
\mbox{e}^{-\frac{(z^2+z^{*\,2})\Sigma^2 V}{8\mu^2F_\pi^2}} 
K_0\left(\frac{|z|^2\Sigma^2 V}
{4\mu^2F_\pi^2}\right)\int_0^1 dt t e^{-2\mu^2F_\pi^2V t^2} I_0(z\Sigma
V)I_0(z^*\Sigma V). \nn \\
\ee
The unquenched eigenvalue density, on the other hand, was derived first by
means of the random matrix techniques \cite{O} and subsequently derived using
the replica trick \cite{AOSV}.  
Here we give the result for one flavor in the topologically trivial sector 
(see \cite{O,AOSV} for the general expressions)   
\be
\rho_{N_f=1}(z,z^*,m;\mu) & = & 
\rho_{N_f=0}(z,z^*;\mu)\left(1-\frac{I_0(z\Sigma V)
\int_0^1 dt t e^{-2\mu^2F_\pi^2V t^2} I_0(m \Sigma V)I_0(z^* \Sigma V)}
{I_0(m\Sigma V)\int_0^1 dt t e^{-2\mu^2F_\pi^2V t^2} I_0(z \Sigma V)I_0(z^* \Sigma V)
}\right).
\label{rhoNf1}
\ee
Note that the unquenched density is the sum of the quenched density and a
term from unquenching $\rho_{N_f=1}(z,z^*,m;\mu)  = 
\rho_{N_f=0}(z,z^*;\mu)+\rho_U(z,z^*,m;\mu)$ and that the density is zero
for $m=z$.

\begin{figure}[!ht]
\begin{center}

\begin{picture}(30,2.0)  
  \put(140,-40.0){\bf\large $y\Sigma V$}
  \put(-225,-20){\bf \LARGE
$\frac{Re[\rho_{N_f=1}(x,y,m;\mu)]}{\Sigma^2V^2}$}  
\put(-225,-90){\bf \LARGE $2\mu= \frac{1}{\sqrt{2}} \, m_\pi$}
  \put(140,-200.0){\bf\large $y\Sigma V$}
  \put(-225,-240.7){\bf \LARGE $2\mu=m_\pi$}
  \put(0,-510){\bf\large $x\Sigma V$}
  \put(140,-394.0){\bf\large $y\Sigma V$}
  \put(-225,-421.7){\bf \LARGE $2\mu= \sqrt{\frac{3}{2}} \, m_\pi$}
\end{picture}

\vspace{0cm}\includegraphics[width=8cm]{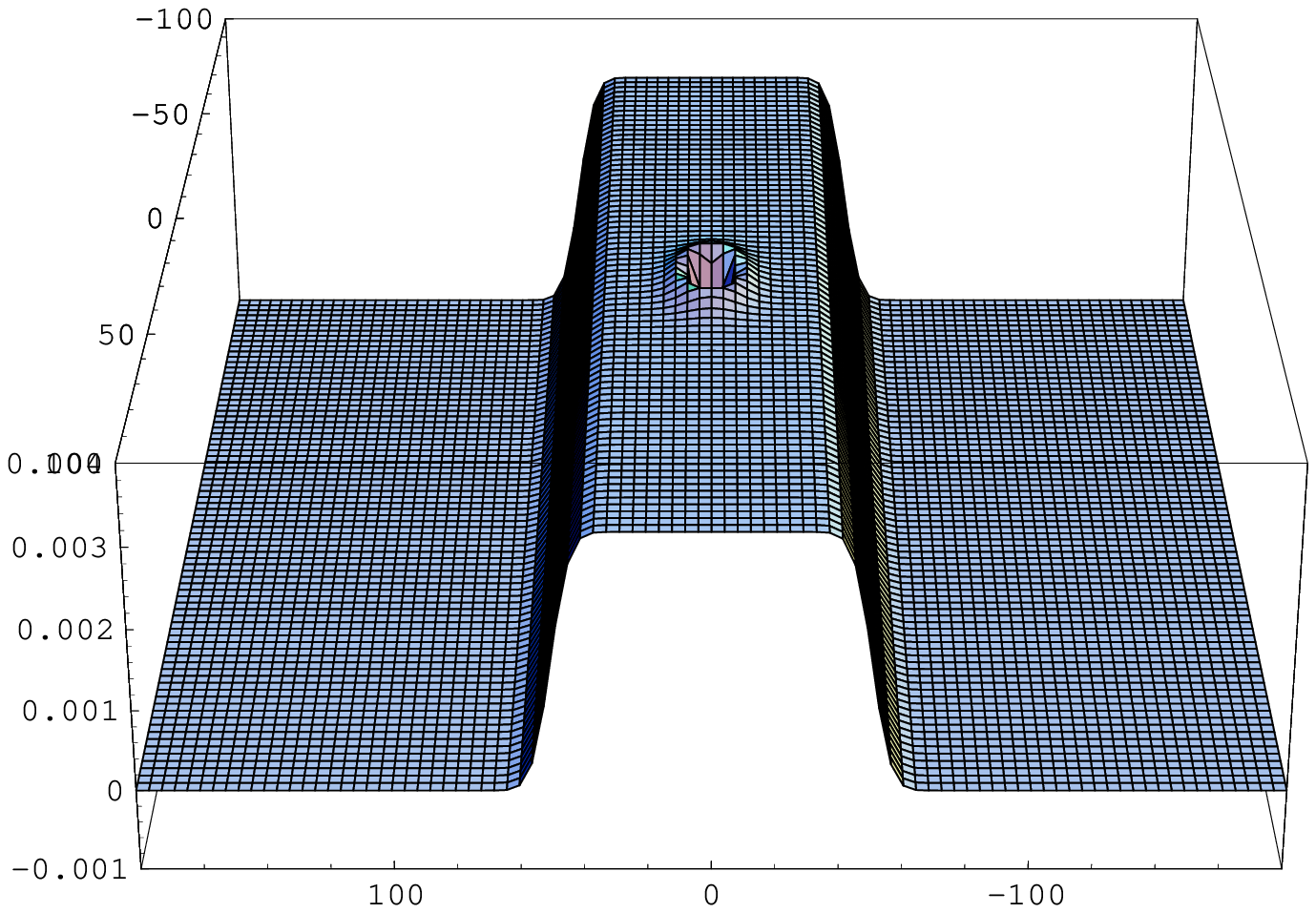}
\vfill
\includegraphics[width=9cm]{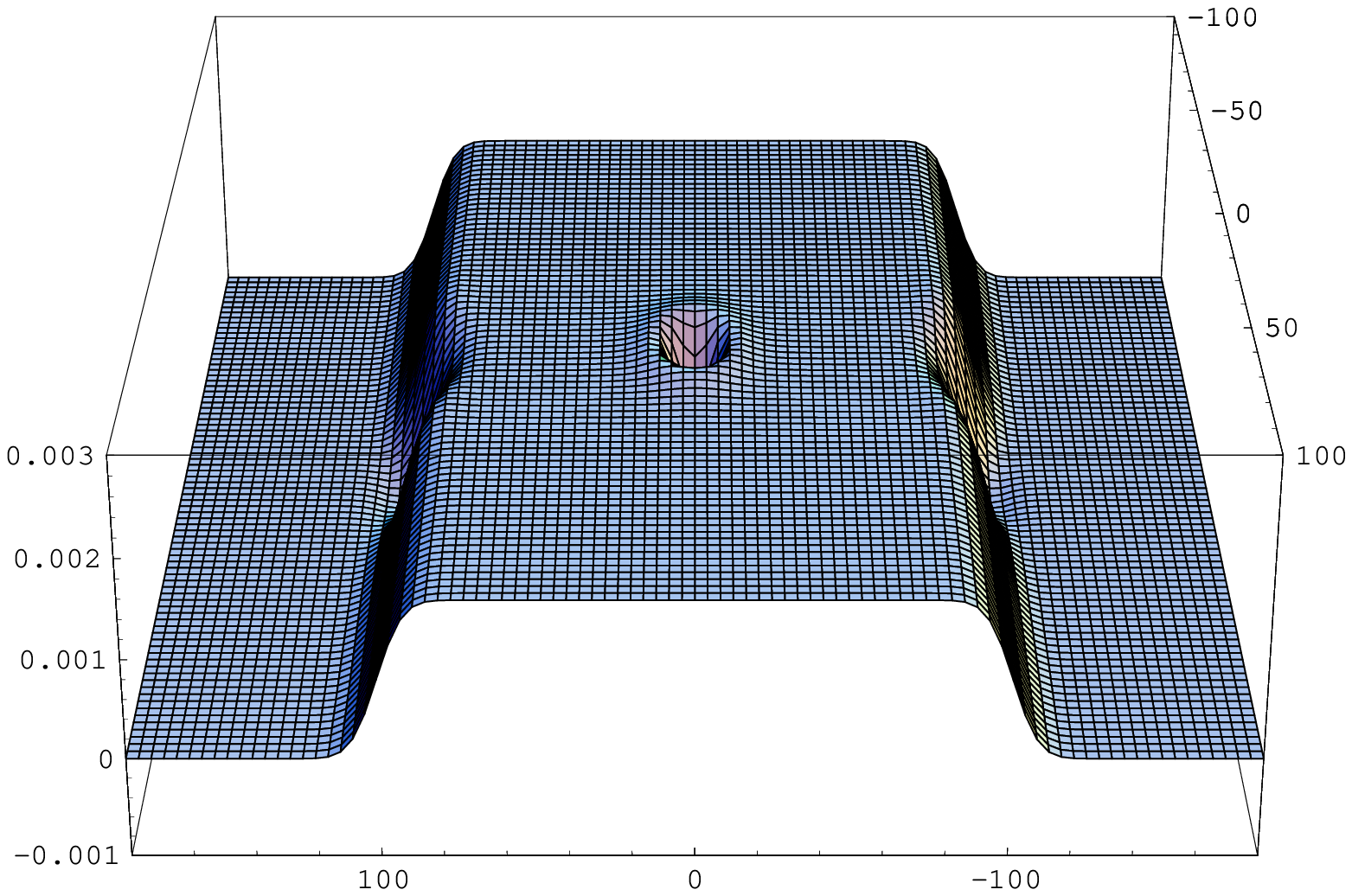}
\vfill
\includegraphics[width=9cm]{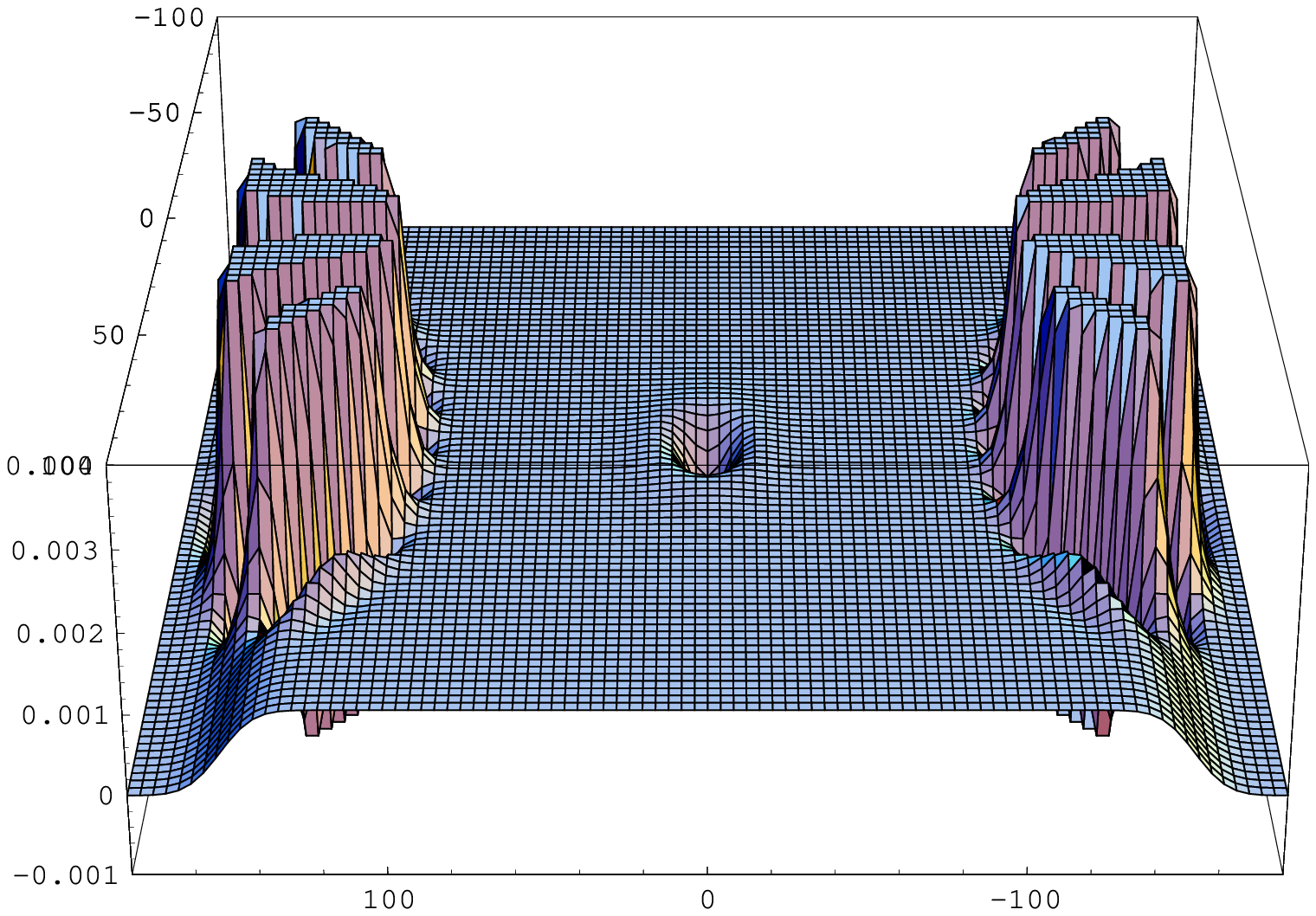}
\vskip -0.3cm
\vskip -0.5cm
\end{center}
\caption{\label{fig:evdens1} The real part of the microscopic eigenvalue density of the QCD
  Dirac operator for fixed quark mass $m\Sigma V=100$. The 
  chemical potential increases from the top and down such that the width 
  $2\mu^2F_\pi^2V=50$, 100, 150. To the left of the plots the chemical
  potential is 
  expressed in terms of the pion mass. Note that the support of the
  eigenvalue distribution reaches the quark mass when
  $\mu=m_\pi/2$. As $\mu$ exceeds this value two oscillating regions
  starts at $z=\pm m$ and extend towards the edge of the support.}
\end{figure}

\begin{figure}[!ht]
\begin{center}

\begin{picture}(30,2.0)  
  \put(140,-40.0){\bf\large $y\Sigma V$}
  \put(-225,-20){\bf \LARGE
$\frac{Re[\rho_{N_f=1}(x,y,m;\mu)]}{\Sigma^2V^2}$}  
\put(-225,-70){\bf \LARGE $m_\pi=\sqrt{\frac{3}{2}} \, 2\mu$}
  \put(140,-220.0){\bf\large $y\Sigma V$}
  \put(-225,-240.7){\bf \LARGE $m_\pi=2\mu$}
  \put(0,-520){\bf\large $x\Sigma V$}
  \put(140,-406.0){\bf\large $y\Sigma V$}
  \put(-225,-421.7){\bf \LARGE $m_\pi= \sqrt{\frac{1}{2}}\, 2\mu$}
\end{picture}

\vspace{0cm}\includegraphics[width=9cm]{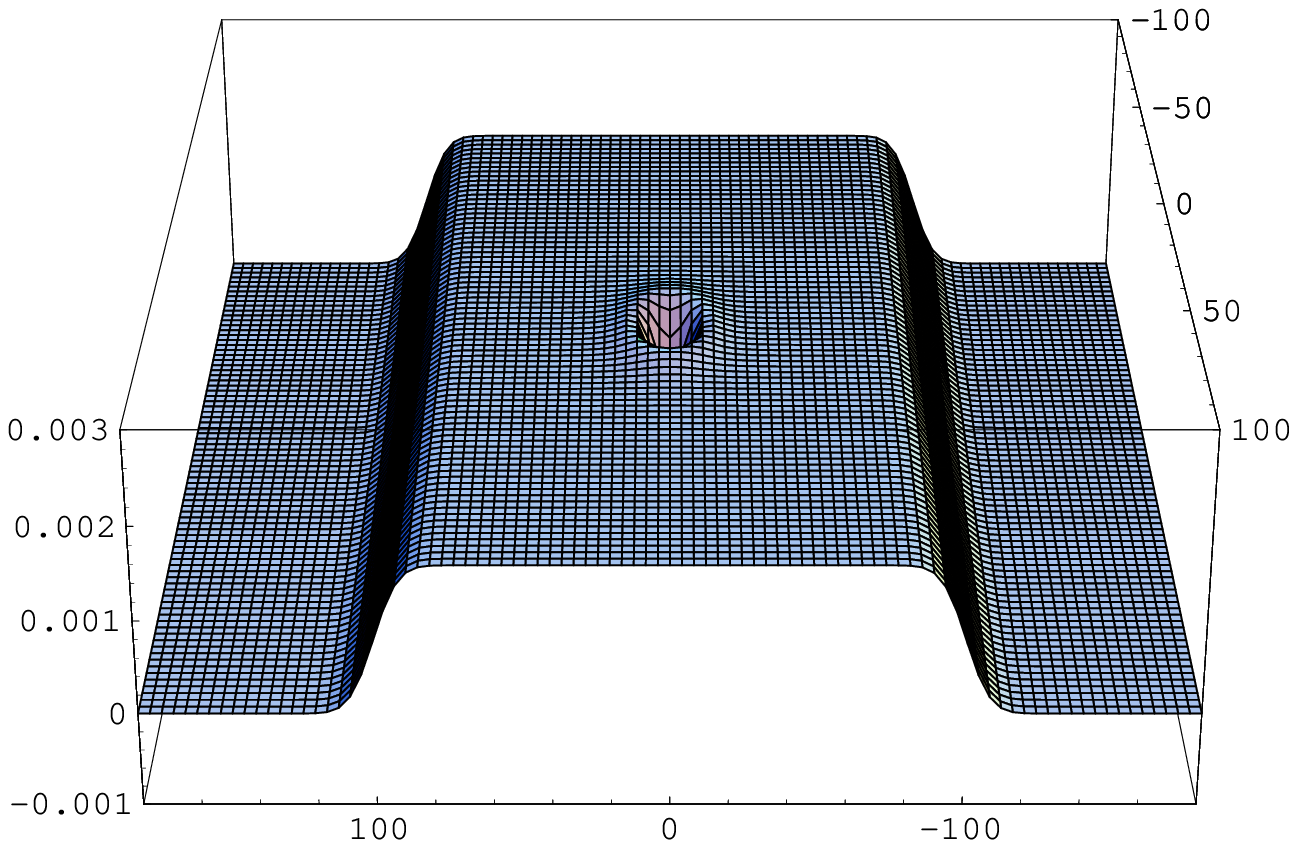}
\vfill
\includegraphics[width=9cm]{m100.eps}
\vfill
\includegraphics[width=9cm]{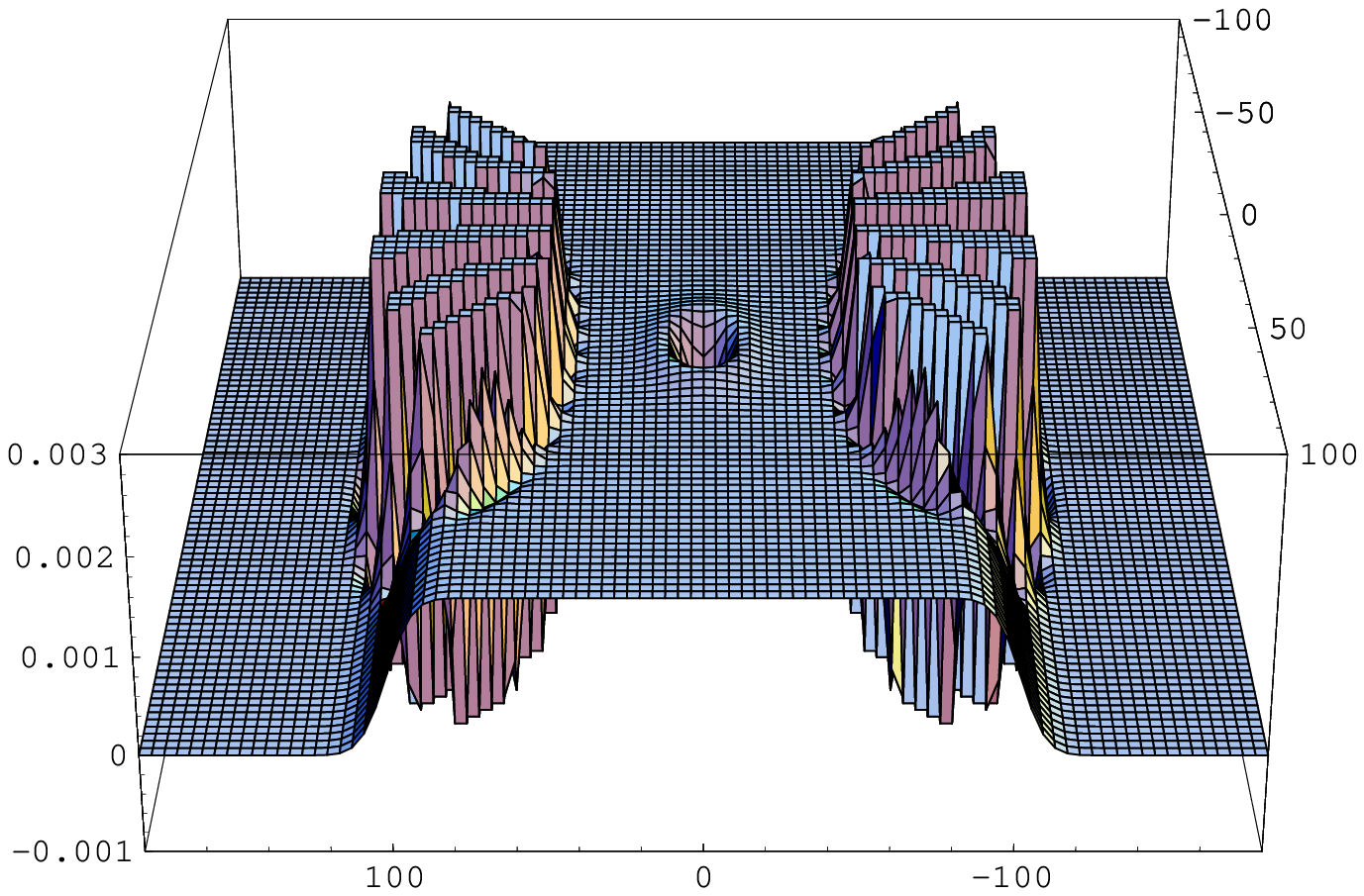}
\vskip -0.3cm
\vskip -0.5cm
\end{center}
\caption{\label{fig:evdens2} The unquenched eigenvalue density for
  fixed chemical potential $2\mu^2F_\pi^2V=100$ and decreasing quark
  mass $m\Sigma V=150$, 100, 50. The pion mass in units of the chemical
  potential is given to the left of the plots. Inside the oscillating
  regions the 
  imaginary part of the eigenvalue density is nonzero and oscillates
  out of phase with the real part shown. Note that the 
  oscillations by far exceeds the scale of the plot.}  
\end{figure}

\section{A complex and oscillating eigenvalue density}

The eigenvalue $z$ and its complex conjugate $z^*$ do not enter symmetrically
in the unquenching part of the eigenvalue density (\ref{rhoNf1}). This 
suggests that the unquenched eigenvalue density is in general a complex
function. Here we take a closer look at the unquenched eigenvalue density of
the QCD Dirac operator and identify the scale at which it becomes complex.

For infinitely large quark mass the unquenching term in (\ref{rhoNf1}) is
suppressed and the eigenvalue density is constant and nonzero in a strip 
along the imaginary axis of width $2\mu^2F_\pi^2/\Sigma$. The quark mass 
enters the strip of eigenvalues when $m<2\mu^2F_\pi^2/\Sigma$ or, 
equivalently, when $m_\pi<2\mu$. After the quark mass has entered this strip 
the unquenched eigenvalue density is dramatically different from the quenched 
eigenvalue density. Starting at $z=\pm m$ and extending to the support of the
eigenvalue density are two regions in which the unquenched eigenvalue density
is complex and oscillating. Figures \ref{fig:evdens1} and
\ref{fig:evdens2} illustrate the appearance of the oscillating regions for
$2\mu>m_\pi$.  
The oscillations have an amplitude which grows exponentially with the volume
and have a period inversely proportional to the volume. 
This structure has a physical origin:
The generating functionals for the eigenvalue density in (\ref{replicadens}) 
are partition functions with additional pairs of
conjugate fermions (for explicit evaluation of these partition functions 
in the $\epsilon$-regime see \cite{SplitVerb2,AFV}). 
The three regions of the unquenched 
eigenvalue density correspond to three phases of the generating functionals 
\cite{OSV,proc,KT}.
The uniform part of the density corresponds to the phase with a condensate of
pions made op of the quarks with masses $z$ and $z^*$. This pion condensate
forms when $2\mu$ exceeds the mass of these pions,
$(2\mu)^2>2Re[z]\Sigma/F_\pi^2$. This why the with of the strip of
eigenvalues is $Re[z]<2\mu^2F_\pi^2/\Sigma$. 
The oscillating region
corresponds to a phase with Bose condensation of pions of squared mass 
$(m+z^*)\Sigma/F_\pi^2$. This condensate dominates for $m<z$ since in this
case the 
squared mass, $(m+z^*)\Sigma/F_\pi^2$, of these pions is smaller than 
the squared mass, $(z+z^*)\Sigma/F_\pi^2$, of the pions made up of the quarks
with masses $z$ and $z^*$. Finally, the region outside the support of the
eigenvalue 
density corresponds to the normal phase, without Bose condensates, of the
generating functionals.  

To see how the structure emerges from (\ref{rhoNf1}) 
let us look at the limit where the width of the eigenvalue support is
large, $2\mu^2F_\pi^2V\gg1$, and where the quark mass and eigenvalue is
inside the support and well away from the origin ($m\Sigma
V\gg1$, $|z|\Sigma V\gg1$). In this case the unquenched eigenvalue density
simplifies to ($z=x+iy$)
\be
\rho_{N_f=1}(x,y,m;\mu) \sim
\frac{1}{4\mu^2F^2V}(1-e^{V\Sigma\left[\frac{\Sigma(x^2-y^2+m^2)}{8\mu^2F_\pi^2}
-\frac{\Sigma x^2}{2\mu^2F_\pi^2}+\frac{\Sigma xm}{4\mu^2F_\pi^2}+x-m\right]}
e^{i\, V\Sigma y\left(1-\frac{\Sigma(x+m)}{4\mu^2F_\pi^2}\right)}).
\label{asymp}
\ee
The term independent of $x$ and $y$ (the $'1'$) 
gives the plateau of the quenched part (since we assumed
that the eigenvalue was inside the strip we do not see the boundary at
$x=2\mu^2F_\pi^2/\Sigma$). The second term gives the effect of unquenching, 
it is exponentially suppressed or enhanced with the volume depending on the 
sign in the square bracket. This gives the boundary of the oscillating
region. Finally, from the oscillating exponential it is clear that the period of the
oscillations is of order $1/V$. For $x$ and $m$ well inside the support the
oscillations are predominantly along the imaginary axis.    

The oscillations are of course a manifestation of the sign problem so we
should expect that the sign problem is acute for $\mu>m_\pi/2$. We will
confirm this expectation in section \ref{sec:strength}.

\section{The Banks-Casher relation at nonzero $\mu$}
\label{sec:BC}

At zero chemical potential the accumulation of eigenvalues at the origin on
the imaginary axis is responsible for chiral symmetry breaking \cite{BC}. 
As we discuss now the oscillations of the eigenvalue density are responsible
for chiral symmetry breaking at $\mu\neq0$ \cite{OSV}. 

\begin{figure}[!ht]
\begin{center}

\begin{picture}(30,2.0)  
  \put(00,-25.0){\bf\large $y\Sigma V$}
  \put(-225,-20){\bf \large
$\frac{Re[\rho_{N_f=1}(x,y,m;\mu)]}{\Sigma^2V^2}$}
  \put(-90,-125){\bf\large $x\Sigma V$}
  \put(-225,-150.7){\bf \large
$\frac{\rho_{N_f=0}(x,y;\mu)}{\Sigma^2V^2}$}
  \put(-225,-270){\bf \large
$\frac{Re[\rho_{U}(x,y,m;\mu)]}{\Sigma^2V^2}$}
\end{picture}

\vspace{1cm}
\includegraphics[width=7cm]{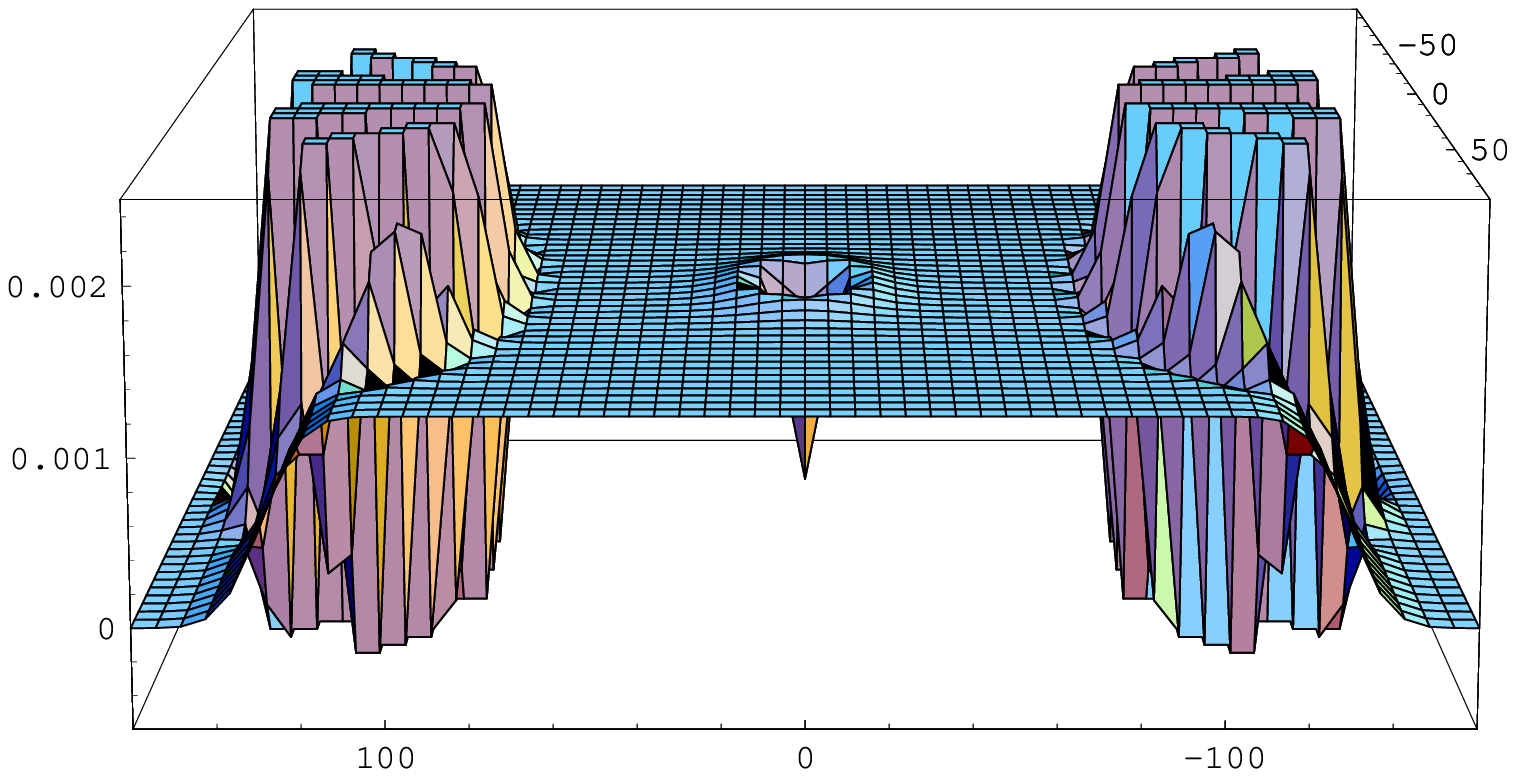}
\includegraphics[width=7cm]{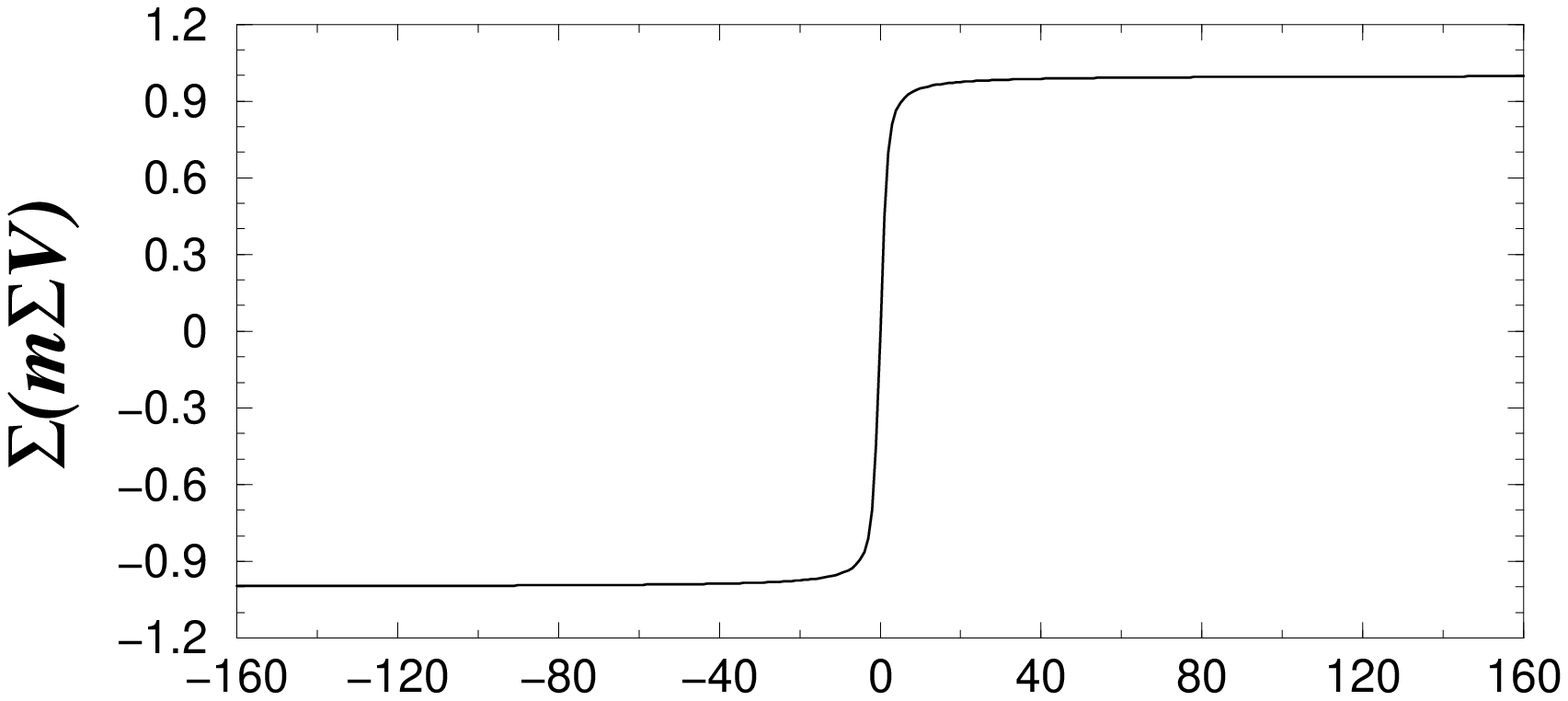}
\vfill
\vspace{5mm}

\includegraphics[width=7cm]{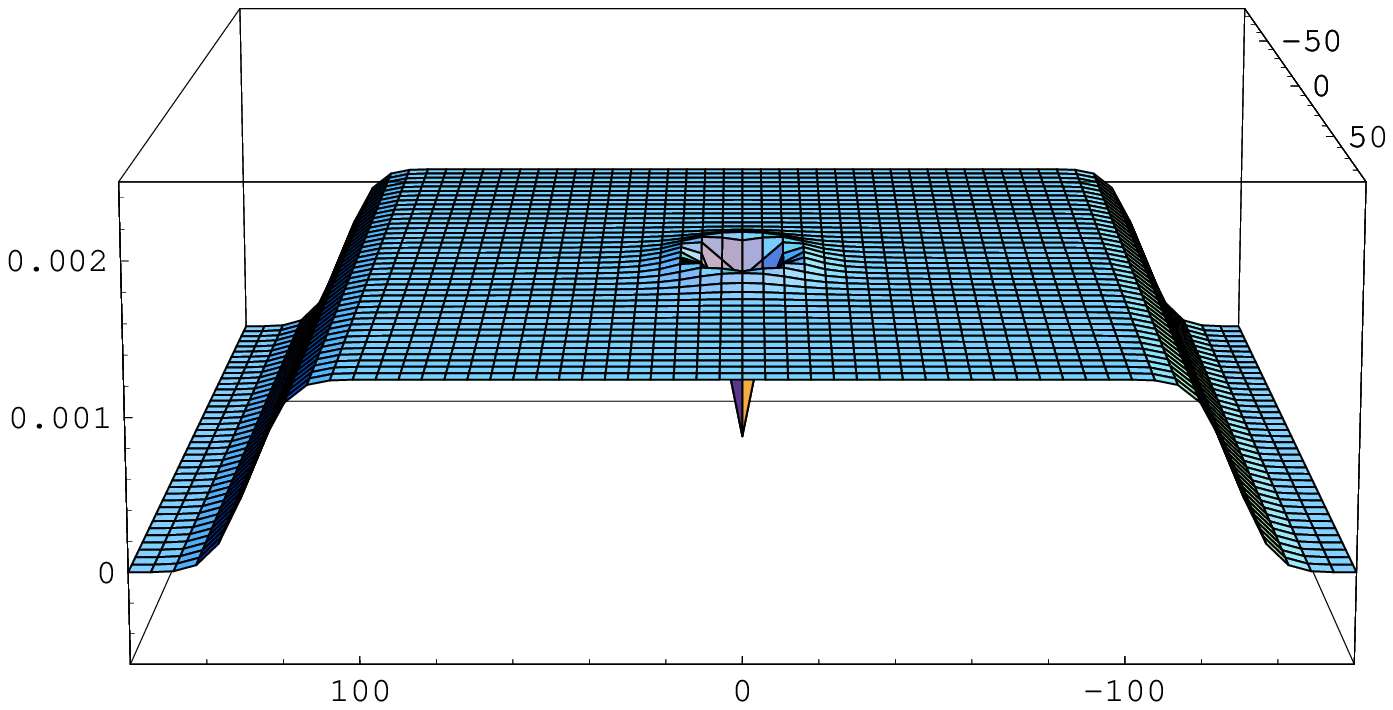}
\includegraphics[width=7cm]{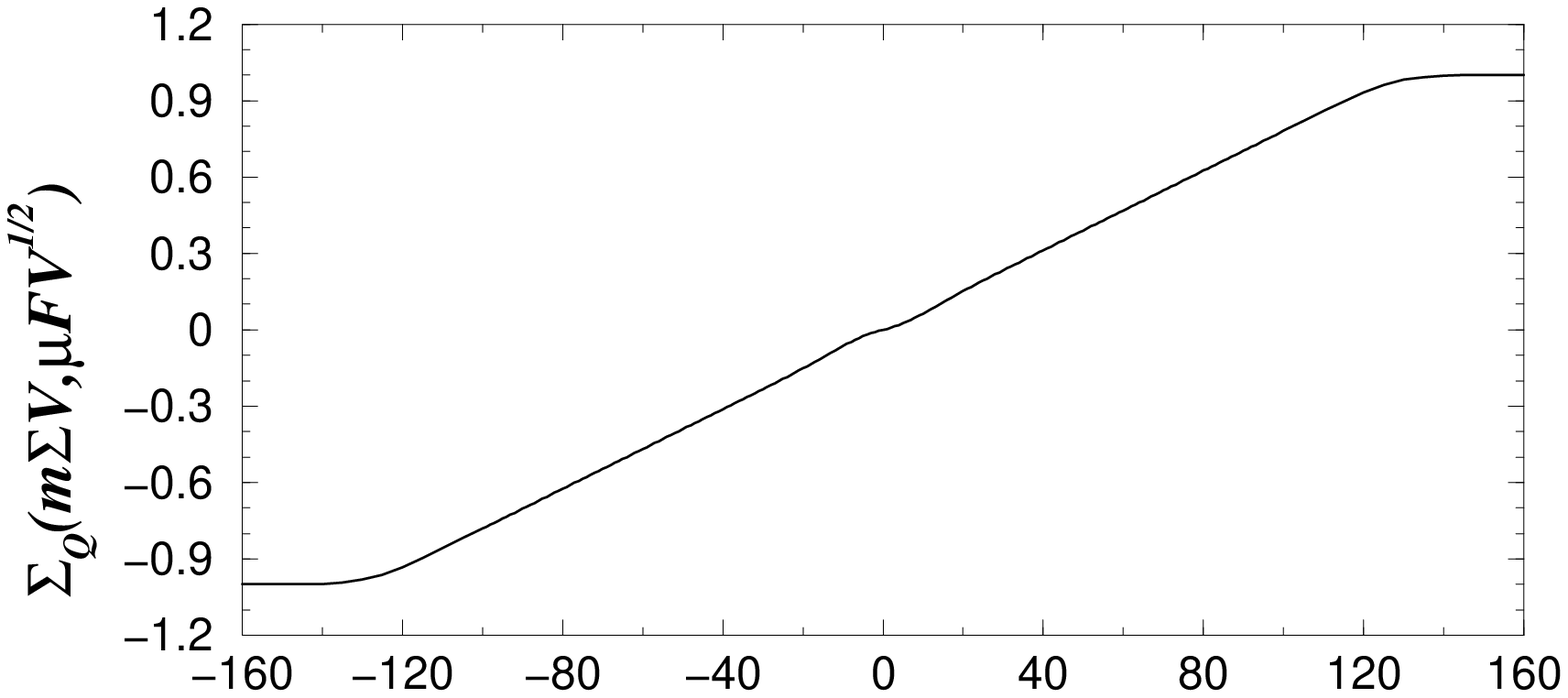}
\vfill
\vspace{5mm}

\includegraphics[width=7cm]{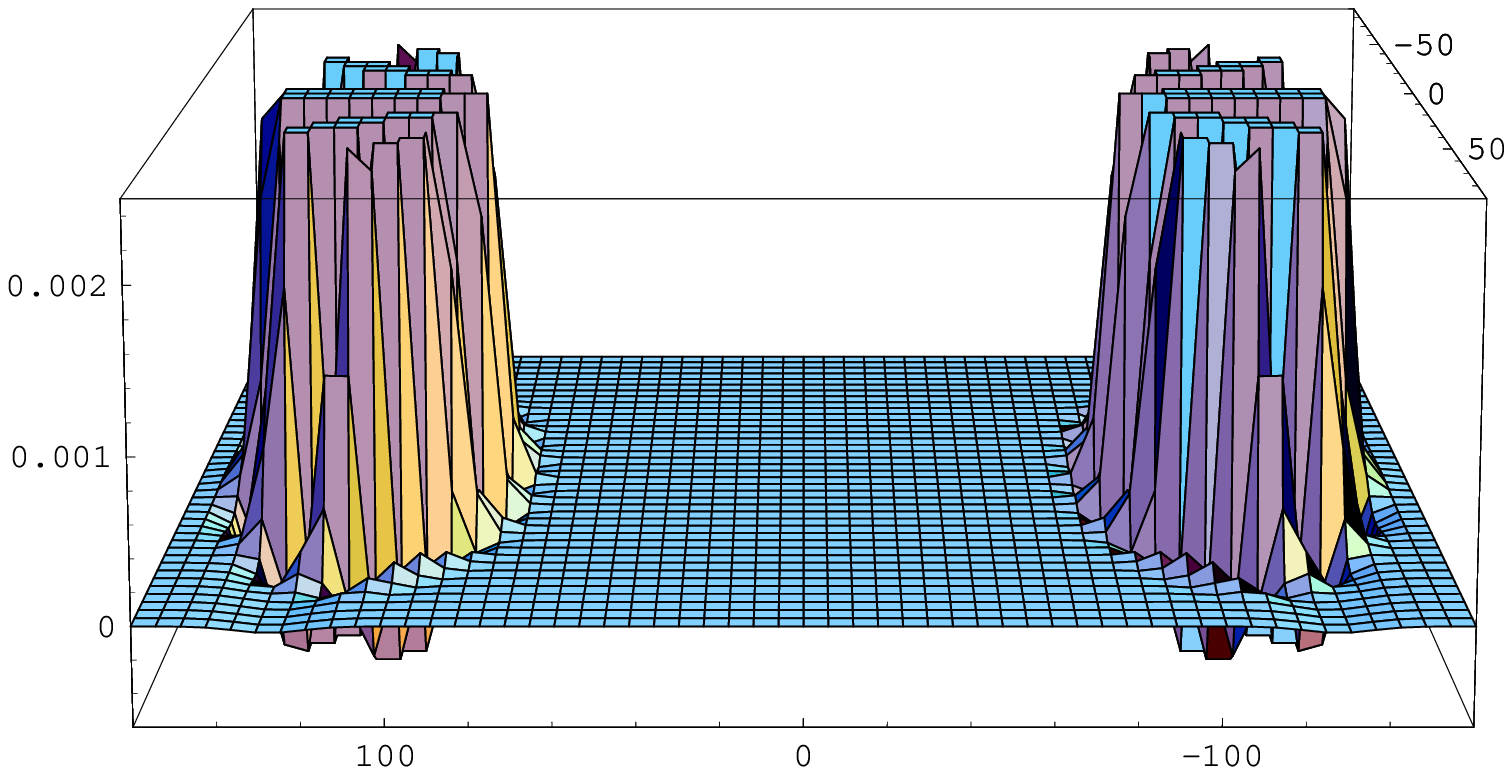}
\includegraphics[width=7cm]{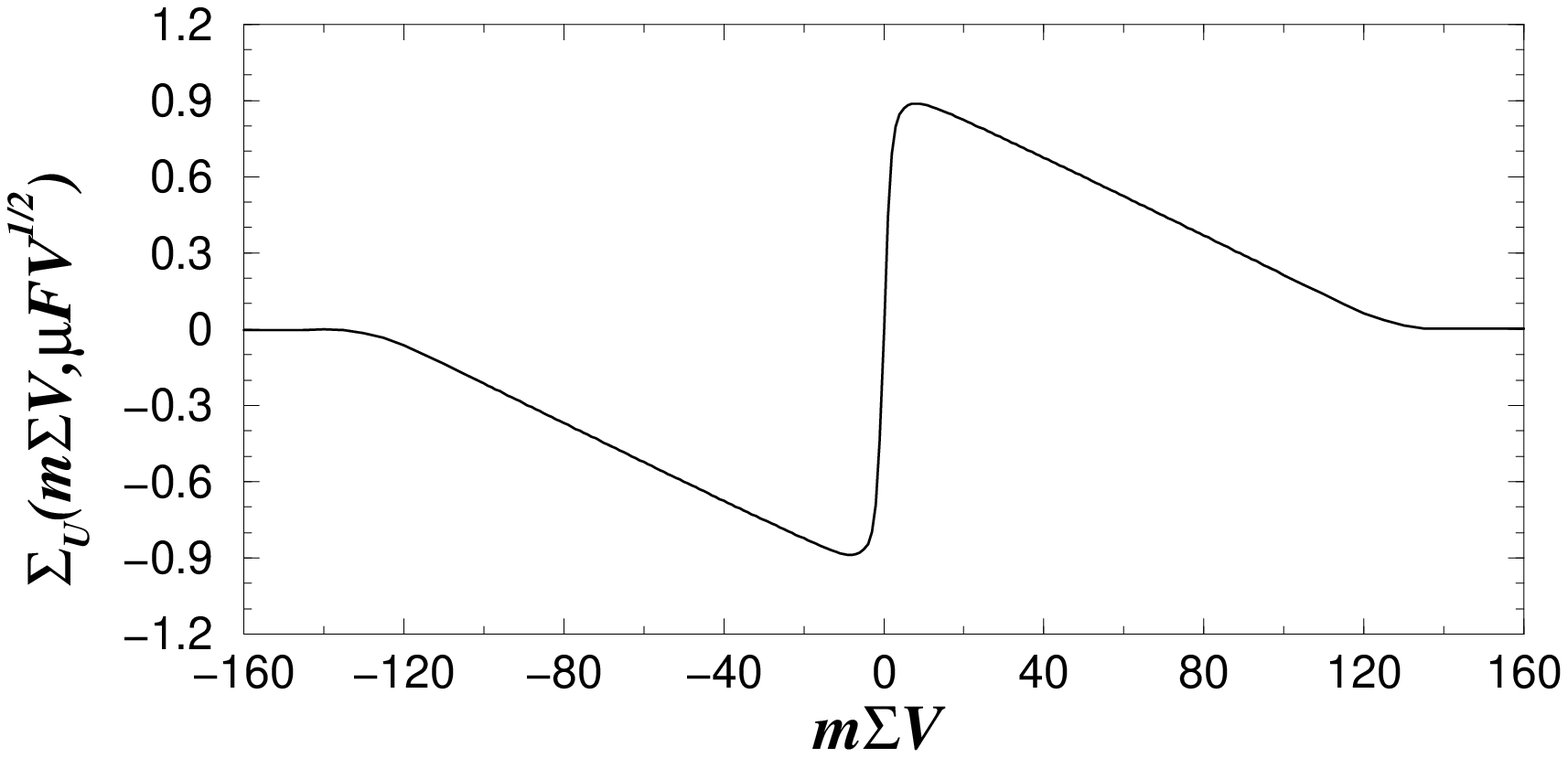}
\vskip -0.3cm
\end{center}
\caption{\label{fig:BCmuneq0}
  {\bf Left column:} The unquenched eigenvalue density (top)
  split into the quenched part (middle) and the oscillating part
  (bottom). {\bf Right column:} The chiral condensate as a function of
  the quark mass. The top panel shows the full chiral condensate and
  the two plots below give the individual contributions from
  the quenched eigenvalue density and the oscillating region
  respectively. Note 
  that it is the oscillations of the eigenvalue density which are
  responsible for the discontinuity of the chiral condensate at zero
  quark mass.} 
\end{figure}

Using the definition of the eigenvalue density (\ref{defdens}) the 
chiral condensate can be expressed as an integral over the complex
eigenvalue plane
\be
\Sigma_{N_f}(m;\mu) = \int d^2z \
\frac{\rho_{N_f}(z,z^*,m;\mu)}{z+m}. 
\label{Sigmafromrho}
\ee
As noted below eq. (\ref{rhoNf1}) is it natural to write the unquenched
eigenvalue density as a sum of two terms
\be
\rho_{N_f}(z,z^*,m;\mu) = \rho_{N_f=0}(z,z^*;\mu) + \rho_{U}(z,z^*,m;\mu),
\label{rhoQ+U}
\ee 
where $\rho_{U}(z,z^*,m;\mu)$ is what is left after subtracting the quenched
density. See the left hand column of figure \ref{fig:BCmuneq0}. 
The quenched density is by definition real and positive so, in
agreement with our asymptotic analysis above, the complex oscillations
reside entirely in $\rho_{U}(z,z^*,m;\mu)$. Inserting 
$\rho_{N_f}=\rho_{N_f=0}+\rho_{U}$ in (\ref{Sigmafromrho}) shows that the
chiral condensate is built up from two terms 
\be
\Sigma_{N_f} =\Sigma_Q+\Sigma_{U}. 
\ee
The individual contributions to the chiral condensate are shown the right
hand column of figure \ref{fig:BCmuneq0}. As expected 
from lattice simulations and the electrostatic analogy the quenched
contribution drops to zero in the chiral limit. The entire discontinuity
of the chiral condensate thus comes from the oscillating part, cf. the lower
row of  figure \ref{fig:BCmuneq0}. To see analytically how the 
oscillations of the eigenvalue density build up the discontinuity,
insert the asymptotic form (\ref{asymp}) in (\ref{Sigmafromrho}) 
and first perform the integral over $y$ for fixed $x$ by going to the complex
$y=a+ib$ plane \cite{OSV}. In this plane the roles of the two exponentials in
(\ref{asymp}) are mixed: Since there is an explicit factor
of $V$ in both arguments, the second exponential now also affects the boundary
of the oscillating region (this is why it is essential that the
oscillations have a period  of order $1/V$ and an amplitude which is
exponentially large in the volume).  
Due to the mixing the contour can be deformed into a region where the
integrand is exponentially suppressed. 
The $y$-integral through the oscillating part is therefore given  
by the residue at the pole alone. The residue follows automatically from the 
observation that the unquenched eigenvalue density vanishes at $z=m$
\cite{OSV}.

\section{Lattice simulations in the $\epsilon$-regime at $\mu\neq0$}

The sign problem in QCD occurs since the baryon chemical potential introduces
a mismatch between quarks and anti-quarks. If we instead consider a
chemical potential for the third component of isospin then the anti-particle
is a part of the measure which therefore remains real \cite{AKW}. 
Quenched QCD with 
$\mu\neq0$ is the zero flavor limit of QCD at nonzero isospin chemical 
potential \cite{misha}. The microscopic eigenvalue density 
(\ref{rhoquenched}) in the quenched case has been compared successfully to
staggered lattice simulations \cite{AW,OW} as well as to simulations of a 
Ginsparg-Wilson Dirac operator at nonzero chemical potential \cite{BW}. 
The measure of 2 color QCD at nonzero baryon chemical potential is also real
and this has allowed to test the predictions for the microscopic
spectral density \cite{A2col} in the quenched case \cite{ABLMP} as
well as the unquenched \cite{AB}. 

If the baryon chemical potential is purely imaginary the Dirac operator
remains anti-hermitian. The correlations between two such Dirac operators
separated by a microscopic difference between the values of the imaginary 
chemical potential is extremely sensitive to the value of $F_\pi$. The
correlation function thus provides a way to extract the pion decay constant
from simulations in the $\epsilon$-regime \cite{DHSS}.

\section{The strength of the sign problem}
\label{sec:strength}

The analysis of the spectra of the QCD Dirac operator has shown  
that the sign problem manifest itself in the eigenvalue density when 
$\mu>m_\pi/2$. This is precisely the value of $\mu$ for which the eigenvalue
density reaches the quark mass and thus where eigenvalues, $z$, for which
$(z-m)\sim 1/\Sigma V$ become frequent. 
In order to quantify the strength of the sign problem let us write 
\be
\det(D+\mu\gamma_0+m) = |\det(D+\mu\gamma_0+m)|e^{i\theta}
\ee
and consider the expectation value of the (squared) phase factor 
of the fermion determinant 
\be
\left\langle e^{2i\theta} \right\rangle_{N_f} =  
\left\langle \frac{\det(D+\mu\gamma_0+m)}{\det(D+\mu\gamma_0+m)^{\Huge\bf *}}
 \right\rangle_{N_f}=\frac{Z_{N_f+1|1^*}}{Z_{N_f}}.
\label{ratio}
\ee 
For $\mu=0$ the phase, $\theta$, is zero and $\langle e^{2i\theta}\rangle=1$.
If the fluctuations drives $\langle e^{2i\theta}\rangle$ to zero
the sign problem is very strong.

\begin{figure}[h] 
\begin{center}
  \includegraphics[width=10cm]{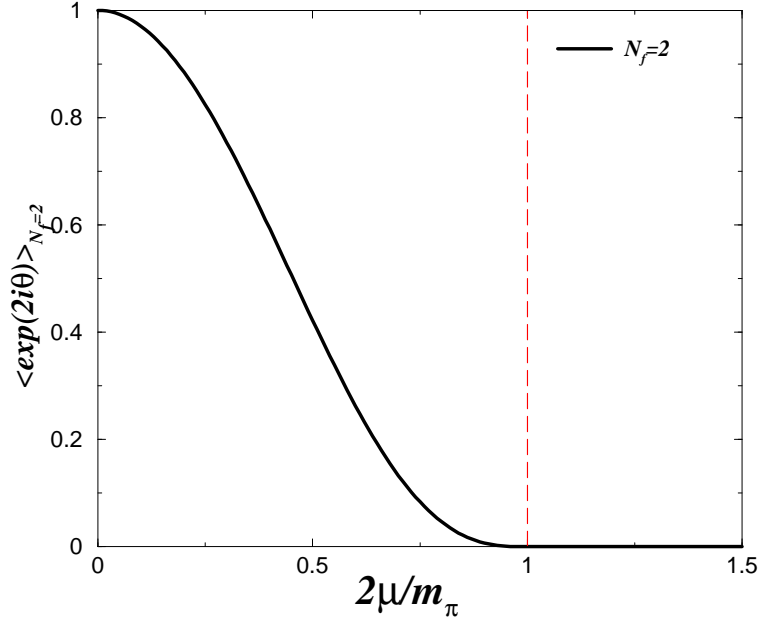}
\end{center}
  \caption{\label{fig:phase}The expectation value of the phase factor of the
    fermion determinant in  
    QCD with two dynamical flavors. Shown is the curve for $\mu F_\pi
    \sqrt{V}\gg1$. In this limit of the $\epsilon$-regime
    $\langle\exp(2i\theta)\rangle$ 
    only depends on the ratio $2\mu/m_\pi$.}
\end{figure}

The expectation value of the phase factor is equal to the partition function 
($Z_{N_f+1|1^*}$) with an additional fermionic flavor and an additional
conjugate bosonic flavor 
divided by the standard dynamical partition function ($Z_{N_f}$). In the
$\epsilon$-regime these partition functions can be evaluated explicitly
\cite{boson,phase}. The thermodynamic limit, $m\Sigma V\to\infty$ and $\mu^2 F_\pi^2
V\to\infty$, of the result is extremely simple. 
The sign problem has two distinct phases (see figure \ref{fig:phase}): 
For $\mu<m_\pi/2$ the sign problem saturates at a nonzero value 
\be
\left\langle e^{2i\theta} \right\rangle_{N_f}=
(1-\frac{4\mu^2}{m_\pi^2})^{N_f+1} e^{V0} \ \ \  {\rm for} \ \ \ \mu<m_\pi/2 .
\label{exp2ith}
\ee
In the other phase the sign problem is exponentially bad in the volume 
\be
\left\langle e^{2i\theta} \right\rangle_{N_f} 
\sim e^{-VF_\pi^2\frac{(m_\pi^2-4\mu^2)^2}{8\mu^2}} \ \ \  {\rm for} \ \ \
\mu>m_\pi/2.
\ee
The behavior of the phase follows directly from leading order chiral
perturbation theory \cite{phase}. At leading order the low energy effective 
partition function is given by a saddle point approximation
\be
Z^{\rm LO} \sim  J\sqrt{\frac{\prod_f V m_{\pi,f}^2}{\prod_b V
    m_{\pi,b}^2}}e^{-V\Omega_{MF}}, 
\label{ZMF}
\ee 
where $m_{\pi,b}$ ($m_{\pi,f}$) are the masses of the Goldstone bosons
(fermions). (The Goldstone fermions are made up of 
a fermionic quark and a bosonic
conjugate quark.) The free energy density $\Omega_{MF}$ is given by the
Lagrangian at the mean field value of the fields. Finally, $J$ is the
Jacobian from the measure.

For $\mu<m_\pi/2$ the additional fermionic and bosonic
quark in the numerator of (\ref{ratio}) has no effect on the mean field free 
energy. The factors of $\exp(-V\Omega_{\rm MF})$ hence cancel between the
numerator and denominator in (\ref{ratio}).  
For $\mu>m_\pi/2$ the partition function in the numerator is in a Bose
condensed phase and, consequently, the mean field free energy in the numerator
does not match that in the denominator. This leads to an expectation value of
the phase which is exponentially small in the volume.

The saddle point approximation, (\ref{ZMF}), also allow us to understand 
the preexponential factor in (\ref{exp2ith}). The masses of the charged 
pions do depend on $\mu$ even for $\mu<m_\pi/2$ \cite{KST,KSTVZ,eff}. 
The charged pions are the ones made up from a fermionic quark
and a bosonic conjugate quark. There are $2(N_f+1)$ such Goldstone fermions, 
half of which have mass $m_\pi-2\mu$ while the other half have mass 
$m_\pi+2\mu$. The resulting overall factor $(m_\pi^2-4\mu^2)^{N_f+1}$ is
divided by $m_\pi^{2(N_f+1)}$ which is left after canceling out the neutral
Goldstone bosons from the partition function in the denominator. This
explains the result (\ref{exp2ith}) for $\langle e^{2i\theta}\rangle_{N_f}$
when $\mu<m_\pi/2$.

Since $\langle e^{2i\theta} \rangle_{N_f}$ is a ratio of two partition
functions it is real. The same is true for $\langle e^{-2i\theta}
\rangle_{N_f}$ implying that 
$\left\langle\sin(2\theta)\right\rangle_{N_f}$ is purely imaginary. To see
that $\left\langle\sin(2\theta)\right\rangle_{N_f}$ is nonzero we need to 
compute $\langle e^{-2i\theta} \rangle_{N_f}$ and show that it is different 
from $\langle e^{2i\theta} \rangle_{N_f}$. The expectation value of the 
inverse phase factor also has a very simple form in the thermodynamic limit
\be
\left\langle e^{-2i\theta} \right\rangle_{N_f} =
(1-\frac{4\mu^2}{m_\pi^2})^{-N_f+1}e^{V0} \ \ \  {\rm for} \ \ \ \mu<m_\pi/2.  
\ee 
The expectation value of the inverse phase is {\sl not} equal to 
the expectation of the phase. They are only equal in quenched and phase
quenched QCD where the weight function is real. 

Note that lattice tests of the predictions for $\langle e^{2i\theta}\rangle$ 
are possible for $\mu<m_\pi/2$ even in unquenched QCD. Related observables 
has been measured on the lattice \cite{Nakamura,Ejiri,BieSwan}.

\section{Nonzero temperature}

Unquenched lattice simulations at nonzero chemical potential must deal with
the sign problem. At present three major approaches have been explored, the
multi parameter reweighting method \cite{FK}, the Taylor expansion method
\cite{BieSwan,GG}, and analytic continuation from imaginary chemical
potential \cite{DeFPh,Lombardo}. See also the plenary review at this lattice
conference by C. Schmidt \cite{Christian}.

\begin{figure}[!ht]
\begin{center}

\begin{picture}(30,2.0)  
  \put(-60,-155){\small $\langle\exp(2i\theta)\rangle=0$}
  \put(-65,-175){\footnotesize Ev density oscillates}
  \put(-70,-115){\footnotesize {\bf \large $\leftarrow$} Quark mass hits} 
  \put(-65,-125){\footnotesize               eigenvalue density}
  \put(152,-135){\footnotesize 4th root illdefined}
  \put(142,-170){\footnotesize Yang-Lee analysis fails}
\end{picture}

\vspace{1cm}
\includegraphics[width=7.5cm]{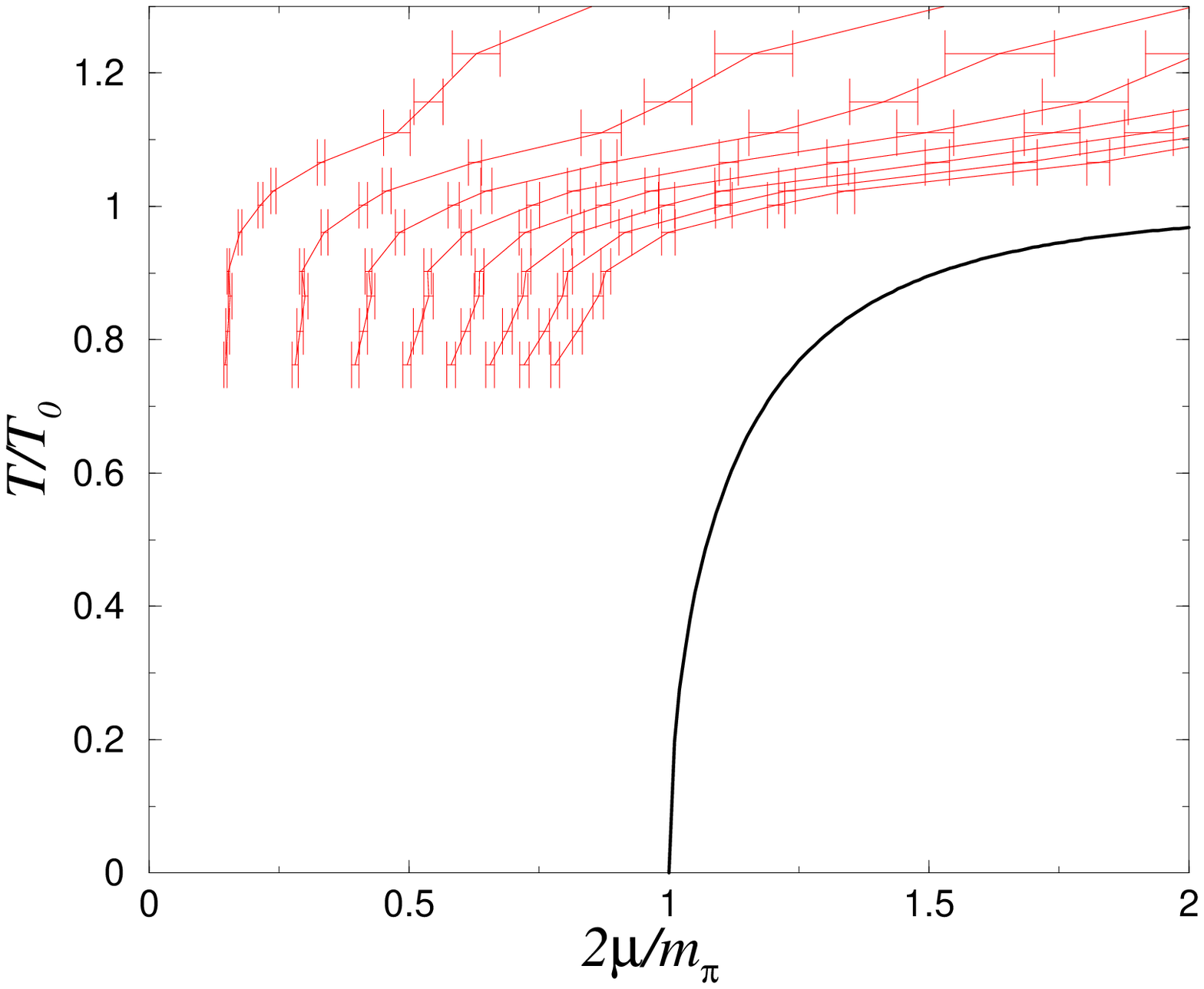}
\includegraphics[width=7.5cm]{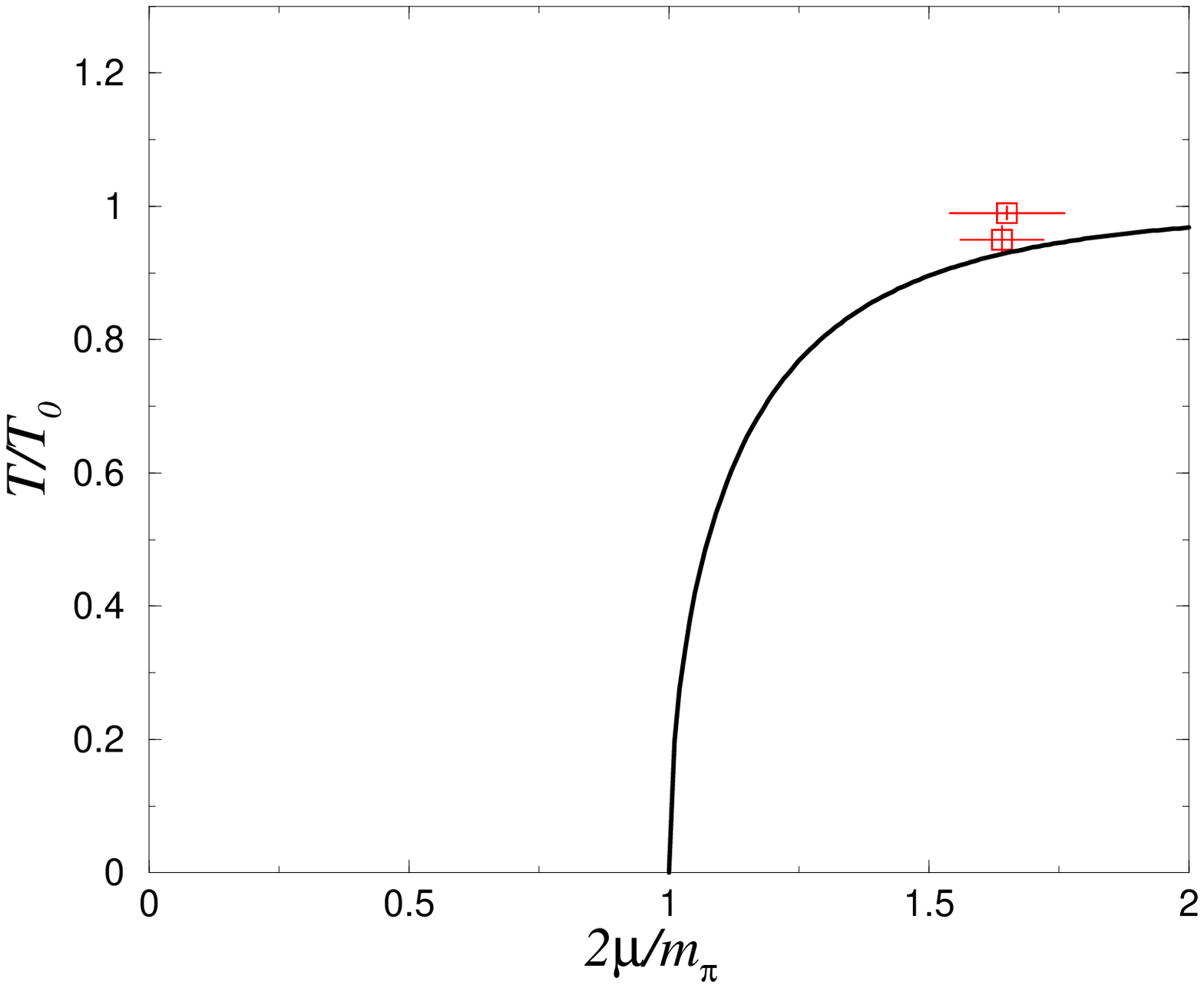}

\end{center}
\caption{\label{fig:Tneq0}{\bf Left:} Contour lines of the variance
  $\sqrt{\langle\theta^2\rangle-\langle\theta\rangle^2}=\pi/4,2\pi/4,\ldots 2\pi$ 
  of the phase of the fermion determinant from \cite{BieSwan}. Note that the
  variance depends on the distance from the line where the quark mass hits
  the support of the eigenvalue spectrum (indicated by the thick line). 
  {\bf Right:} The measured endpoints from \cite{FK} and the line where the 
  quark mass hits the eigenvalue density. To the right of this line the
  Yang-Lee analysis \cite{Ejiri} and the 4th root trick \cite{GSS} used in 
  \cite{FK} are troublesome.}    
\end{figure}

Above we computed the strength of the sign problem and found that it changes
drastically when $\mu=m_\pi/2$. The fact that the change takes place at
$\mu=m_\pi/2$ has a physical origin: $\langle\exp(2i\theta)\rangle$ is 
a ratio of two partition functions, c.f. (\ref{ratio}), and since the 
partition function in the numerator includes a conjugate quark it 
experiences a phase transition into a Bose condensed phase at this 
value of $\mu$. The Bose condensate changes the functional form of the 
free energy and causes an exponential suppression of
$\langle\exp(2i\theta)\rangle$ for $\mu>m_\pi/2$. The region in the 
$\mu,T$-plane for which this Bose condensate is present is therefore 
identical to the region where $\langle\exp(2i\theta)\rangle$ is 
exponentially suppressed. As we have argued above the chemical potential 
above which the quark mass is inside the support of the eigenvalue density is
also determined by Bose condensation of weakly interacting pions. The region 
of the $\mu,T$-plane where the sign problem is exponentially suppressed is
therefore identical to the region where the quark mass is inside the support
of the eigenvalue density. 
This region is below the thick line indicated in figure
\ref{fig:Tneq0}. As the temperature is increased the line bends to the 
right since for a sufficiently high temperature the Bose condensate will
melt. The curve gives the melting temperature of the Bose condensate as
calculated in \cite{KTV}. It is consistent with the melting temperature
computed in phase quenched lattice QCD \cite{KogutS}.

The lattice computations of \cite{BieSwan} suggests that the strength
of the sign problem scales with the critical chemical potential:
In the left panel of figure \ref{fig:Tneq0} we show contour lines from
\cite{BieSwan} of the variance of the two flavor staggered fermion 
determinant, in our notation
$\sqrt{\langle(2\theta)^2\rangle-\langle2\theta\rangle^2}$. 
The lines give the contours up to the value $2\pi$ in steps of
$\pi/4$. The contours are parallel to the thick line indicating the  
expected critical chemical potential for which the quark mass hits the
support of the eigenvalue density and the sign problem becomes very severe 
\cite{muBmuI}.

Almost all of the lattice simulations \cite{FK,BieSwan,GG,DeFPh,Lombardo}
address the region of the $\mu,T$-plane where the sign problem is less severe. 
One exception is \cite{FK}. These two studies of the critical endpoint used
two sets of quark masses. The endpoint was found to depend strongly
on the quark mass. In fact, as observed in \cite{GG}, the value of the chemical
potential at the endpoint scales like pion mass in these studies. In the
right hand part of figure \ref{fig:Tneq0} we give the location of the
endpoints found in \cite{FK} together with the line where the quark
mass is expected to hit the support of the eigenvalue density. Surprisingly
both endpoints points are located very close to the line \cite{muBmuI}. 
To the right of the
line the sign problem is exponentially strong and this has been argued
\cite{Ejiri} to invalidate the Yang-Lee analysis used in \cite{FK}. 
Moreover, when the quark mass is inside the support of the eigenvalue 
density the 4th root trick used in \cite{FK} may be illdefined \cite{GSS}. 
For these reasons one
may fear that the signals interpreted as a signature of the endpoint
in \cite{FK} is rather a breakdown of the method used.

\section{Summary}

The study of the $\epsilon$-regime of QCD at nonzero chemical potential has
provided new insights in QCD which go beyond the microscopic scale. 
Here we have reviewed those aspects which have direct relevance for the sign
problem in lattice QCD.  
The non-hermitian nature of the Dirac operator at $\mu\neq0$ also have a very
nontrivial effect on the analytical methods which where used to derive the 
results discussed here. For a review focused on these aspects, see
\cite{gernot-review}. 

Here we have discussed how the analysis of the spectrum of the QCD Dirac
operator allows us to understand the sign problem in lattice QCD at nonzero
chemical potential. From the perspective of the Dirac operator the sign
problem plays an all important role for spontaneous chiral symmetry breaking.
The sign problem induces violent complex oscillations in the spectral density
of the Dirac operator and these in turn build up the entire discontinuity of
the chiral condensate in the chiral limit. Therefore, to address spontaneous
chiral symmetry breaking in the chiral limit on the lattice at $\mu\neq0$, 
one must deal with the sign problem. 

The strength of the sign problem can be measured by the average of the phase
factor, $\langle\exp(2i\theta)\rangle$. In general
$\langle\exp(2i\theta)\rangle$ depends on quark mass, the 
chemical potential, the volume, the temperature as well as the lattice
cutoff. In the 
$\epsilon$-regime we can quantify the dependence of
$\langle\exp(2i\theta)\rangle$ on $\mu$ and the quark mass: 
 $\langle\exp(2i\theta)\rangle$ is nonzero for $\mu<m_\pi/2$ while for
$\mu>m_\pi/2$ it is exponentially small in the volume. 
The separation between these two scales is linked to the onset 
of Bose condensation and this physical insight allows us to
extrapolate the results beyond the $\epsilon$-regime. For $\mu>m_\pi/2$ we
expect that there is a critical temperature at which the sign problem changes
its nature. Care should be taken not to misinterpret manifestations of
this change in lattice QCD.

\vspace{1cm}

\noindent
{\bf Acknowledgments:} 
It is a privilege to thank Jac Verbaarschot, Gernot Akemann, James Osborn, Dominique
Toublan, Poul Henrik Damgaard, Urs Heller and Benjamin Svetitsky for
collaborations. My warmest tanks also to the organizers of lattice 2006
for creating a perfect work environment. The author is supported by the
Carlsberg Foundation.

\end{document}